\documentstyle[preprint,aps,pra,amssymb]{revtex}
\tighten 
\draft 
\begin{document} 
\title{Ultra-low energy scattering of a He atom  off a He dimer} 
\author{E. A. Kolganova} 
\address{Laboratory of Computing Techniques and Automation, 
	Joint Institute  for Nuclear Research, Dubna, 141980, Russia} 
\author{ A. K. Motovilov\thanks{On leave of absence from 
	 the Laboratory of Theoretical Physics, 
    Joint Institute for Nuclear Research, Dubna, 141980, Russia}, 
	S. A. Sofianos} 
\address{Physics Department, University of South Africa,  P.O.Box 
	  392, Pretoria 0001, South Africa} 
\maketitle 
 
\begin{abstract} 
We present a new, mathematically rigorous, method suitable for 
bound state and scattering processes calculations for various 
three atomic or molecular systems where the underlying forces 
are of a hard-core nature. We employed this method to 
calculate the binding energies and the ultra-low energy 
scattering phase shifts below as well as above the break-up 
threshold for the three He-atom system. The method is proved to 
be highly successful and suitable for solving the three-body 
bound state and scattering problem in configuration space and 
thus it  paves the way  to study various three-atomic systems, 
and to calculate important quantities such as the 
cross-sections, recombination rates etc. 
 
\medskip
 
    LANL E-print {\tt physics/9802016}.

    Published in Phys. Rev. A., 1997, v.~56, No.~3, pp.~1686--1689R.

\end{abstract} 
\bigskip 
 
 
The $^4$He trimer is of interest in various areas of Physical 
Chemistry and Molecular Physics, in particular such as the 
behavior of atomic clusters under collisions and Bose-Einstein 
condensation. Various theoretical and experimental works have 
been devoted in the past to study its ground state properties 
and in general the properties of the $^4$He and other noble gas 
droplets. From the theoretical works we mention here those using 
Variational and Monte Carlo type methods~\cite{V1,V2,V3,V4,V5}, 
the Faddeev equations~\cite{Nakai,Gloeckle,CGM}, and the 
hyperspherical approach~\cite{Levinger,Sof,EsryLinGreene}.  From 
the experimental works we recall those of 
Refs.~\cite{ArgonExp,XenonExp,DimerExp,Science} where molecular 
clusters consisting of a small number of noble gas atoms were 
investigated. 
 
Despite the efforts made to solve the He-trimer problem various 
questions such as the existence of Efimov states and the study 
of scattering processes at ultra-low energies still have not 
been satisfactorily addressed. In particular  for scattering 
processes there are no works which we are aware of apart from a 
recent study concerning recombination rates~\cite{Fed96}. There 
are various reasons for this the main one being the fact that 
the three-body calculations involved are extremely difficult to 
perform due to the practically hard--core of the inter--atomic 
potential which gives rise to strong numerical inaccuracies that 
make calculations for these molecules cumbersome and unstable. 
 
In the present work we employed a hard core version of the 
Boundary Condition Model (BCM)~\cite{EfiSch,MMYa}, developed 
in~\cite{MerMot}, \cite{Vestnik}. The so-called hard-core 
potentials represent a particular variant of the BCM where one 
requires that the wave function vanishes when particles approach 
each other at a certain distance $r=c$. Such a requirement is 
equivalent to an introduction of an infinitely strong repulsion 
between particles at distances $r\leq c$. The standard formalism 
for scattering (see, for example,~\cite{MF}) does not deal with 
hard-core interactions described by these boundary conditions. 
Replacement of the finite, for $r>0$, but often singular at 
$r=0$, repulsive short-range part of the potential with a 
hard-core interaction turns out to be a very effective way to 
suppress inaccuracies related to a numerical approximation of 
the Schr\"odinger operator at short distances. 
 
In order to outline our method we start from the Schr\"odinger 
equation for bound states, 
$ 
	      H\Psi=E\Psi\,, 
$ 
where $\Psi$ is the three-body bound state wave function.  We 
are concerned with states for which $E<0$ and that these 
energies are below the threshold of the  continuous spectrum of 
$H$. Using the Green's formula~\cite{Potential} one can show 
that the function $\Psi$ satisfies the following 
Lippmann-Schwinger type equation 
\begin{equation} 
\label{LSchCor} 
     \Psi(X)=-\displaystyle\int\limits_{\partial\Omega} 
     d\sigma_S\, G_0(X,S;E)\frac{\partial}{\partial n_S}\Psi(S) 
     -\displaystyle\sum_{\alpha=1}^3\, 
     \displaystyle\int\limits_{\Omega} dX'\, G_0(X,X';E) 
    (V_{\alpha}\Psi)(X')\,. 
\end{equation} 
Here  $\Omega$ is the configuration space of the three-body 
system in the hard-core model which represents only a part of 
the six-dimensional space, ${{\Bbb R}}^{6}$, external, $|{\bf 
x}_{\alpha}|> c_{\alpha}$, with respect to all three cylinders 
$\Gamma_\alpha$, 
$ 
        \Gamma_{\alpha}=\{X\!\in\!{{\Bbb R}}^{6}:\, 
	X=({\bf x}_{\alpha}, {\bf y}_{\alpha}),\, 
	|{\bf x}_{\alpha}|=c_{\alpha}\} 
$, 
$\alpha=1,2,3$, where $c_{\alpha}$, $c_{\alpha}>0$, stands for 
the value of $|{\bf x}_{\alpha}|$ when the cores of the 
particles in the pair $\alpha$ contact each other. The 
${\bf x}_{\alpha}, {\bf y}_{\alpha}$ are the usual Jacobi 
coordinates~\cite{MF}. By $G_0(X,X';z)$ we denote the free 
Green function of the three-body Laplacian $-\Delta_X$ and by 
$n_S$, the external unit vector (directed into $\Omega$) normal 
to the surface $\partial\Omega$ while $d\sigma_S$ is a surface 
element (five-dimensional square) on $\partial\Omega$. By 
$V_\alpha$, $V_\alpha=V_\alpha({\bf x}_\alpha)$, we denote 
the pair potentials acting outside the core domains, i.e. at $|{\bf 
x}_\alpha|>c_\alpha$. 
 
The Faddeev components of the function $\Psi$ are introduced via 
the formulas \cite{MMYa,MerMot} 
\begin{equation} 
\label{FaddComp} 
       \Phi_{\alpha}(X)=-\displaystyle\int\limits_{\Gamma_{\alpha} 
       \bigcap\partial\Omega}d\sigma_S\, G_0(X,S;E) 
       \frac{\partial}{\partial n_S} \Psi(S) 
       - \displaystyle\int\limits_{\Omega} dX'\, G_0(X,X';E)\, 
       ( V_\alpha\Psi)(X')\,. 
\end{equation} 
One can show that they satisfy the following system 
of differential equations 
\begin{equation} 
\label{FaddeevEq} 
\left\{ 
    \begin{array}{rcll} 
        (-\Delta_X+V_{\alpha}-E)\Phi_{\alpha}(X) 
        &=&-V_{\alpha} \displaystyle\sum 
        \limits_{\beta\neq\alpha}\Phi_{\beta}(X)\,, 
        \, \, & \,\, |{\bf x}_{\alpha}|>c_{\alpha}\,, \\ 
        (-\Delta_X-E)\Phi_{\alpha}(X) 
         &=& 0\,, \,\, & \,\, |{\bf x}_{\alpha}|<c_{\alpha}\,. 
     \end{array}\right. 
\end{equation} 
According to Eqs.~(\ref{LSchCor}) and (\ref{FaddComp}) the sum 
$\Phi_{\alpha}(X)$ outside the surface 
$\partial\Omega$ coincides with the total wave function $\Psi$, 
i.e., 
$ 
   \sum_{\beta=1}^3 \Phi_{\beta}(X)\equiv 
   \Psi(X), \quad X\in{\Omega}\,. 
$ 
At the same time, it follows from the Green's formula that this 
sum vanishes inside all the core domains, 
$ 
   \sum_{\beta=1}^3 \Phi_{\beta}(X)\equiv 
   0, \quad X\in{{\Bbb R}}^6\setminus{\Omega}\,. 
$ 
In practice  one can replace these, very strong, conditions 
with the essentially more weak 
ones~\cite{MerMot,Vestnik} 
\begin{equation} 
\label{SummFaddCylinder} 
       \left. \displaystyle\sum_{\beta=1}^3 \Phi_\beta(X)\right |_{ 
       |{\bf x}_\alpha|=c_\alpha}=0, \qquad \alpha=1,2,3\,, 
\end{equation} 
requiring that the sum of $\Phi_{\alpha}(X)$ to be zero only on 
the cylinders $\Gamma_{\alpha}$. 
 
Partial version of the Faddeev equations~(\ref{FaddeevEq}) for a system 
of three identical bosons read (see~\cite{MF,MGL}) 
\begin{equation} 
\label{FadPart} 
   \left(-\displaystyle\frac{\partial^2}{\partial x^2} 
            -\displaystyle\frac{\partial^2}{\partial y^2} 
            +\displaystyle\frac{l(l+1)}{x^2} 
            +\displaystyle\frac{\lambda(\lambda+1)}{y^2} 
    -E\right)\Phi_{aL}(x,y)=\left\{ 
            \begin{array}{cl} -V(x)\Psi_{aL}(x,y), & x>c \\ 
                    0,                  & x<c 
\end{array}\right. 
\end{equation} 
Here, by $x,y$ we denote absolute values of the Jacobi variables 
$\bf x,y$ and by $c$, the core size which is now the same for 
all three two-body subsystems. The notation $L$ stands for the 
total angular momentum, and $l,$ $\lambda$, for the relative 
angular momenta corresponding respectively to a two-body 
subsystem and complementary particle; $a=\{l,\lambda\}$.  The 
potential $V(x)$ is supposed to be central, acting in the same 
way in all the partial waves $l$.  The partial wave function 
$\Psi_{aL}(x,y)$ is related to the partial Faddeev components 
$\Phi_{aL}(x,y)$ by 
\begin{equation} 
\label{FTconn} 
         \Psi_{aL}(x,y)=\Phi_{aL}(x,y) + \sum_{a'}\int_{-1}^{+1} 
         d\eta\,h_{a a'}^L(x,y,\eta)\,\Phi_{a'L}(x',y') 
\end{equation} 
where 
$$ 
          x'=\sqrt{\displaystyle\frac{1}{4}\,x^2+\displaystyle 
    \frac{3}{4}\,y^2-\displaystyle\frac{\sqrt{3}}{2}\,xy\eta}\,, 
\qquad 
         y'=\sqrt{\displaystyle\frac{3}{4}\,x^2+\displaystyle 
   \frac{1}{4}\,y^2+ \displaystyle\frac{\sqrt{3}}{2}\,xy\eta}\,, 
$$ 
with $\eta=\hat{\bf x}\cdot\hat{\bf y}$. The explicit form for 
the function $h_{aa'}^L$ can be found in Ref.~\cite{MF,MGL}. 
The functions $\Phi_{aL}(x,y)$ satisfy the boundary conditions 
\begin{equation} 
\label{BCStandard} 
      \Phi_{aL}(x,y)\left.\right|_{x=0} 
      =\Phi_{aL}(x,y)\left.\right|_{y=0}=0\,, 
\end{equation} 
while the partial version of the hard-core 
condition~(\ref{SummFaddCylinder}) is given by 
\begin{equation} 
\label{BCCorePart} 
       \Phi_{aL}(c,y) + \sum_{a'}\int_{-1}^{+1} 
       du\,h_{a a'}^L(c,y,\eta)\,\Phi_{a'L}(x',y')=0\,. 
\end{equation} 
For the  bound-state problem one requires that the functions 
$\Phi_{aL}(x,y)$ are square integrable in the quadrant $x\geq 0$, 
$y\geq 0$. A more detailed and useful in bound state 
calculations is the asymptotic condition 
\begin{eqnarray} 
        \Phi_{aL} &=& \sum\limits_\nu \psi_{l,\nu}(x)\, 
	h_\lambda(\sqrt{E-\epsilon_{l,\nu}}\,y) 
	\left[{\rm a}_{aL,\nu}+o(y^{-1/2})\right]\nonumber\\ 
\label{BScond} 
	&+& \frac{\exp({\rm i}\sqrt{E}\rho+i\pi L/2)}{\sqrt{\rho}} 
	\left[A_{aL}(\theta)+o(\rho^{-1/2})\right] 
\end{eqnarray} 
where $E$ is the bound-state energy, 
$\rho=|X|$, $\theta=\arctan{y/x}$, and $\psi_{l,\nu}(x)$ is the 
two-body partial wave function corresponding to a $\nu$-th bound 
state $\epsilon_{l,\nu}$ for the angular momentum value $l$. 
The notation $h_{\lambda}$ is used for the spherical Hankel 
function.  The coefficients ${\rm a}_{aL,\nu}$ and 
$A_{aL}(\theta)$ describe contributions into $\Phi_{aL}$ from 
the  $(2+1)$ and $(1+1+1)$ channels respectively. The 
corresponding asymptotic boundary conditions for the partial 
Faddeev components of the $(2+1\rightarrow 2+1\,;\,1+1+1)$ 
scattering wave function as $\rho\rightarrow\infty$ and/or 
$y\rightarrow\infty$ read as 
\begin{eqnarray} 
       \Phi_{a'L}^{[a,\nu]}(x,y,p) & = & 
       \delta_{a'a}\psi_{l,\nu}(x)j_\lambda(py) \nonumber\\ 
       &+& \displaystyle\sum\limits_{\nu'} \psi_{l',\nu'}(x) 
       h_{\lambda'}(\sqrt{E-\epsilon_{l',\nu'}}\,y) 
       \left[{\rm a}_{a'L,\nu'}^{[a,\nu]}(p)+o\left(y^{-1/2} 
       \right)\right] 
\label{AsBCPart} \\ 
 &+& \displaystyle\frac{\exp({\rm i}\sqrt{E}\rho+ 
     {\rm i}\pi L/2)}{\sqrt{\rho}} 
     \left[A_{a'L}^{[a,\nu]}(p,\theta)+ 
     o\left(\rho^{-1/2}\right)\right] \nonumber
\end{eqnarray} 
where $p=|{\bf p}|$ is the relative moment conjugate to the 
variable $y$ and the scattering energy $E$ is given by 
$E=\epsilon_{l,\nu}+p^2$. The  $j_{\lambda'}$ stands for the 
spherical Bessel function. The value ${\rm 
a}^{[a,\nu]}_{a'L,\nu'}$ represents, at $E> \epsilon_{l',\nu'}$, 
the partial amplitude of an elastic scattering, $a'=a$ and 
$\nu'=\nu$, or rearrangement,  $a'\neq a$ or $\nu'\neq\nu$, 
process.  The functions $A_{a'L}^{[a,\nu]}(\theta)$ provide us, 
at $E>0$, the corresponding partial Faddeev breakup amplitudes. 
 
We employed the Faddeev equations~(\ref{FadPart}) and the 
hard-core,~(\ref{BCCorePart}), and 
asymptotic,~(\ref{BScond},\ref{AsBCPart}), boundary conditions 
to calculate the binding energies of the Helium atomic trimer 
and the ultra--low energy phase shifts of the Helium atom 
scattered off the Helium diatomic molecule. In our calculations 
we take $\hbar^2/m=12.12$\,K\,\AA$^2$.  As a $^4$He--$^4$He 
interatomic interactions we use the HFDHE2~\cite{Aziz79} and 
HFD-B~\cite{Aziz87} potentials of Aziz and co-workers which, we 
found, that they sustain a dimer bound state at $-0.8301$\,mK 
and $-1.6854$\,mK respectively.  At the same time the $^4$He 
atom--$^4$He atom scattering length was found to be 124.7\,{\AA} 
for the HFDHE2 and 88.6\,{\AA} for the HFD-B potential.  

In the present work we restrict ourselves to calculations for 
$S$-state only. The partial components $\Phi_{l\lambda 0}$  can 
be obtained in this case from the addition of even  partial 
waves $l$ and $\lambda$ with  $l=\lambda$. The results of the 
Helium trimer ground-state energy  calculations are presented in 
Table~I. Although the two  potentials used  differ only 
slightly, they produce  important differences in the 
ground-state  energy. This is in agreement with the finding of 
Ref.~\cite{Sof} but in disagreement with the statement made in 
Ref.~\cite{V5}.  It should be  further noted that most of the 
contribution to the binding energy stems from the  $l=\lambda=0$ 
and $l=\lambda=2$ partial component the  latter being more than 
35\,\%. The contribution from the $l=\lambda=4$ partial wave was 
shown in~\cite{CGM} to be of the order of a few per cent. We 
have found that the Helium trimer can form an excited state with 
both the HFDHE2 and HFD-B potentials in agreement with the 
findings of Refs.~\cite{Nakai,Gloeckle,EsryLinGreene}.  Note 
that in the papers~\cite{Gloeckle,EsryLinGreene} this state is 
interpreted as an Efimov one~\cite{VEfimov}. Our excited state 
results are given in Table~II. 
 
The phase shift $\delta_0(E)$ results, for a Helium atom 
scattered off a Helium dimer at $L=0$, are given in Table~III. 
We considered incident energies below as well as above the 
breakup threshold, i.e., for the ($2+1\longrightarrow 2+1$) and 
the ($2+1\longrightarrow 1+1+1$) processes. It is seen that, 
similarly to the bound state results, the inclusion of the 
$l=\lambda=2$ partial wave is essential to describe the 
scattering correctly.  
 
Our estimation for the scattering length, based on the phase 
shift results, with the HFD-B  interactions is 
$170{\pm}5$\,{\AA} in the case where only the  
$l=\lambda=0$ are taken into account  and $145{\pm}5$\,{\AA} 
when both the $l=\lambda=0$ and $l=\lambda=2$ are considered.  
We mention here that an estimation of $\ell_{\rm sc}=195$\,{\AA} 
for the $^4$He atom- $^4$He dimer scattering  
length was previously made by Nakaichi-Maeda and Lim \cite{Nakai}
via zero-energy scattering calculations and by employing
a separable approximation for the HFDHE2 potential.
 
The results obtained with two realistic $^4$He--$^4$He 
potentials clearly demonstrate the reliability of our method in 
three-body bound state and scattering calculations. The 
effectively hard-core inter-atomic potential together with other 
characteristics of the system, make such calculations extremely 
tedious and numerically unstable. The numerical advantage of 
our approach is already obvious from the structure of 
Eqs.~(\ref{FaddeevEq}): When a potential with a strong 
repulsive core is replaced with the hard-core model, one 
approximates, inside the core domains, only the Laplacian 
$-\Delta_X$, instead of the sum of the Laplacian and a huge 
repulsive term, and in this way a much better numerical 
approximation can be achieved. Thus the present formalism paves 
the way to study various three-atomic systems, and to calculate 
important quantities such as the cross-sections, recombination 
rates etc. 
 
 
\bigskip 
\acknowledgements 
Financial support from the University of South Africa, the Joint 
Institute for Nuclear Research, Dubna, and the Russian 
Foundation for Basic Research (Projects No.~96-01-01292, 
No.~96-01-01716 and No.~96-02-17021) is  gratefully 
acknowledged.  The authors are indebted to Dr.~F.~M.~Penkov for 
a number of useful remarks and to Prof.~I.~E.~Lagaris for 
allowing us to use the computer facilities of the University of 
Ioannina, Greece, to perform scattering  calculations.



 
\begin{table} 
\label{tableI} 
\caption 
{Bound state energy (in K) results for the Helium trimer.} 
\begin{tabular}{|c|ccccc|cc|c|} 
\hline 
Potential & \multicolumn{5}{c|}{Faddeev equations} 
&\multicolumn{2}{c|}{Variational}&\multicolumn{1}{c|}{Adiabatic} \\ 
 & \multicolumn{5}{c|}{} 
&\multicolumn{2}{c|}{methods}&\multicolumn{1}{c|}{approach} \\ 
\cline{2-9} 
   & $l$ &  This work & \cite{CGM} & \cite{Gloeckle} 
& \cite{Nakai} & \cite{V2} & \cite{V5} 
& \multicolumn{1}{c|}{\cite{EsryLinGreene}}\\ 
\hline \hline 
HFDHE2 & 0 & 0.084 &  & 0.082 & 0.092 &  &   & 0.098 
\\ \cline{2-6} 
 & 0,2 &  0.114 & 0.107 & 0.11 &  & 0.1173 &  &  \\ 
\hline \hline 
 HFD-B & 0 &0.096 & 0.096 & & & & & \\ 
\cline{2-6} 
 & 0,2 & 0.131  & 0.130 & &  & & 0.1193 & \\ 
\hline 
\end{tabular} 
\end{table} 
 
 
\begin{table} 
\label{tableII} 
\caption 
{Excited state energy (in mK) results for the Helium trimer.} 
\begin{tabular}{|c|c|c|c|c|c|} 
\hline 
Potential   & $l$ &  This work &  \cite{Gloeckle} & \cite{Nakai} 
					  &\cite{EsryLinGreene}\\ 
\hline \hline 
HFDHE2      &  0  &    1.54     &       1.46     &  1.04 & 1.517  \\ 
\cline{2-5} 
            & 0,2 &    1.74     &       1.6      &       & \\ 
\hline \hline 
 HFD-B      &  0  &    2.56     &                &       &\\ 
\cline{2-5} 
            & 0,2 &    2.83     &                &       &\\ 
\hline 
\end{tabular} 
\end{table}

 
\begin{table} 
\label{tableIII} 
\caption 
{The $S$-state Helium atom -- Helium dimer scattering phase shifts 
$\delta_0^{(0)}$ and $\delta_0^{(0,2)}$ obtained with the HFD-B 
$^4$He--$^4$He potential. The shifts $\delta_0^{(0)}$ correspond 
to the case where only the partial wave $l=\lambda=0$ was 
included while the shifts $\delta_0^{(0,2)}$ were obtained with 
inclusion of both partial waves $l=\lambda=0$ and $l=\lambda=2$. 
The values of $\delta_0^{(0)}$, $\delta_0^{(0,2)}$ are given in 
degrees and $E$, in mK.} 
\begin{tabular}{|ccc|ccc|ccc|} 
\hline 
 $E$   & $\delta_0^{(0)}$  & $\delta_{0}^{(0,2)}$  & 
 $E$   & $\delta_0^{(0)}$  & $\delta_{0}^{(0,2)}$  & 
 $E$   & $\delta_0^{(0)}$  & $\delta_{0}^{(0,2)}$  \\ 
\hline 
$-1.685 $& 177.5 & 177.8 & $-1.1$ & 123.6 & 132.6 & 0.7 & 88.9 & 98.7  \\ 
$-1.68$  & 172.3 & 173.3 & $-0.8$ & 115.0 & 124.6 & 1.0 & 85.7 & 95.4  \\ 
$-1.60$  & 153.4 & 156.4 & $-0.4$ & 105.7 & 115.5 & 1.4 & 81.8 & 91.5  \\ 
$-1.5 $  & 142.3 & 148.0 & $-0.1$ &  99.9 & 109.7 & 1.8 & 78.4 & 88.0  \\ 
$-1.4 $  & 135.8 & 143.0 & $+0.3$ &  93.9 & 103.7 & 2.4 & 74.2 & 83.5  \\ 
\hline 
\end{tabular} 
\end{table} 
 
\end{document}